\documentclass[aps,pra,superscriptaddress,twocolumn]{revtex4-1}
\usepackage{bm} \usepackage{graphicx} \usepackage{amsmath}
\usepackage{amssymb} 
\usepackage{braket,mleftright}
\usepackage{fancyhdr}
\usepackage[caption=false]{subfig}
\begin{document}

\title{Tolerance in the Ramsey interference of a trapped nanodiamond}

\author{C. Wan}
\affiliation{QOLS, Blackett Laboratory, Imperial College London, London SW7 2BW, United Kingdom}

\author{M. Scala}\email{m.scala@imperial.ac.uk}
\affiliation{QOLS, Blackett Laboratory, Imperial College London, London SW7 2BW, United Kingdom}

\author{S. Bose}
\affiliation{Department of Physics and Astronomy, University College
London, Gower St., London WC1E 6BT, United Kingdom}
\author{A. C. Frangeskou}
\affiliation{Department of Physics, University of Warwick, Gibbet Hill Road, Coventry CV4 7AL, United Kingdom}

\author{ATM  A. Rahman}
\affiliation{Department of Physics and Astronomy, University College
London, Gower St., London WC1E 6BT, United Kingdom}
\affiliation{Department of Physics, University of Warwick, Gibbet Hill Road, Coventry CV4 7AL, United Kingdom}

\author{G. W. Morley}
\affiliation{Department of Physics, University of Warwick, Gibbet Hill Road, Coventry CV4 7AL, United Kingdom}

\author{P. F. Barker}
\affiliation{Department of Physics and Astronomy, University College
London, Gower St., London WC1E 6BT, United Kingdom}

\author{M. S. Kim}
\affiliation{QOLS, Blackett Laboratory, Imperial College London, London SW7 2BW, United Kingdom}

\date{\today}

\begin{abstract}
The scheme recently proposed in [M. Scala {\em et al.}, Phys Rev Lett {\bf 111}, 180403 (2013)], where a gravity-dependent phase shift is induced on the spin of a nitrogen-vacancy (NV) center in a trapped nanodiamond by the interaction between its magnetic moment and the quantized motion of the particle, provides a way to detect spatial quantum superpositions by means of  spin measurements only. Here, the effect of unwanted coupling with other motional degrees of freedom is considered and we show that it does not affect the validity of the scheme. Both this coupling and the additional error source due to misalignment between the quantization axis of the NV center spin and the trapping axis are shown not to change the qualitative behavior of the system, so that a proof-of-principle experiment can be neatly performed. Our analysis, which shows that the scheme retains the important features of not requiring ground state cooling and of being resistant to thermal fluctuations, can be useful for the several schemes which have been proposed recently for testing macroscopic superpositions in trapped microsystems.
\end{abstract}

\maketitle

\section{Introduction}

Since its discovery more than one century ago, quantum mechanics has puzzled the scientific community with questions foundational. Indeed, while everybody agrees on the power of quantum mechanics for the description of microscopic systems, such as atoms and molecules, and on its applicative power, from electronics to the most recent developments in quantum information processing and computing \cite{NielsenChuang}, there is still a lot of debate about the transition from the microscopic to the macroscopic world, where experience shows that quantum mechanics does not seem to be valid and is substituted by classical physics \cite{zurek2003decoherence}. As an example, the na\"ive use of quantum mechanics for the description of the macroscopic world would lead to predictions contradicting our experience, such as in the  well known case of Schr\"odinger's cat \cite{schrodinger1935gegenwartige}. While it is a relatively simple task to define what the microscopic and the macroscopic worlds are, it is difficult to draw a line separating these worlds. Finding the border between the two worlds is an important foundational issue and, so far, many solutions have been proposed to describe the transition from quantum to classical physics, from more widely accepted theories based on decoherence processes due to the interaction with an external environment \cite{caldeira1983quantum,zurek2003decoherence}, to more debated theories, which propose modifications of quantum mechanics but have not been tested so far, such as the spontaneous localization theories \cite{ghirardi1986unified,pearle1989combining,penrose1996gravity,bassi2013models}

In this framework, mesoscopic physics, i.e. the physics of systems which lie somewhere in between the microscopic and macroscopic worlds, plays an important role, since one can play with parameters which are intuitively related to the transition from quantum to classical, such as the total mass or the total number of atoms involved in the dynamics of the systems. Examples of  experiments with mesoscopic systems are given by double-slit interference with very large molecules \cite{gerlich2011quantum}, the production of nonclassical states of light by means of optomechanical systems \cite{bose1999scheme,rabl2009strong,armour2002entanglement,marshall2003towards} and the study of coherence in trapped nanoparticles \cite{barker2010cavity,yin2013large,romero2010toward,chang2010cavity,romero2012quantum}. In all these experiments, the scientific community is producing quantum superpositions of states of larger and larger systems, and there has been a lot of work about quantifying the {\em macroscopicity} of such superpositions \cite{lee2011quantification,frowis2012measures,nimmrichter2013macroscopicity}. The generally accepted idea is that, by taking more macroscopic regimes and by making the interaction with the environment weaker and weaker, we can finally reach a regime wherein alternative theories, as opposed to orthodox quantum mechanics, will become experimentally testable \cite{bassi2013models}. 
%
%

Usually, the experiments to test such macroscopic superpositions can be quite involved, since in general they require the possibility of cooling the system, coupling it with cavity resonators,  and also resolving the spatial extension of the particles under study \cite{romero2011large}. Moreover in general one will need an ensemble of identically prepared systems. In our previous work  we proposed a quantum interference scheme for trapped nanoparticles which overcomes these requirements and would allow for the detection of the quantum features of the motion of the particle by spin measurements only \cite{scala2013matter}. In the proposal, we considered a conditional displacement induced by a magnetic field gradient on a trapped nanodiamond containing one nitrogen-vacancy (NV) center. Starting from a superposition of two distinct states of the spin of the NV center and from a generic coherent state for the motion of center of the mass (CM) of the trapped bead, we showed that the phase difference acquired by the different trajectories of the harmonic motion in the presence of gravity can be completely transfered to the spin states, so that standard Ramsey interferometry is all we need to detect the gravity-induced phase difference. 

The analysis performed neglected some terms in the Hamiltonian making it effectively one-dimensional, instead of genuinely three-dimensional as it would be in real experiments with optical tweezers. The simplified model allowed us to catch the features of the scheme proposed in an easily understandable way. In this paper we will show that a perturbative approach can be used to prove that those additional terms can be neglected up to very high values of the relevant coupling constants, which implies that the confinement along $x$ and $y$ need not be as tight as one would naively think: this is due to the fact that the first order correction to the energy due to the additional terms are zero and so one has to go to second order to get corrections to the dynamics. We will indeed show that, for parameters describing correctly the dynamics of a nanodiamond trapped in an optical tweezer, the fidelity between the  perturbatively corrected state and the state predicted by the one-dimensional model is always larger than $99 \%$. The same perturbative techniques can be used to treat the effect of misalignment between the quantization axis of the spin of the NV center, which essentially depends on the relative direction between the color center and the diamond lattice, and the trapping axis $z$. We will show that misalignment does not change the qualitative behaviour of the system dynamics and that we are still able to detect interference fringes showing the gravitationally-induced phase difference between the spin states. Moreover, the most important feature of the one-dimensional scheme, i.e., the robustness of the phase difference against thermal fluctuations, is not substantially affected by the additional terms in the Hamiltonian: therefore the more realistic scheme shows that no (or a very small amount of) cooling is required in order to get the interference fringes, so that the simplicity of the proposal remains.
%
%
The paper is structured as follows. In Section II we review the one-dimensional scheme, showing how the quantized motion of the particle leaves phase-signatures on the states of the spin of the NV center, while the effect of the coupling with the motion along the $x$ and $y$ directions and the effect of misalignment are presented in Sections III and IV respectively. In Section V the results are discussed and some concluding remarks are given.
\section{Gravitational-induced phases on spin states}
\subsection{The system and its dynamics}
\begin{figure}
\begin{center}
\includegraphics[width=2.0in, clip]{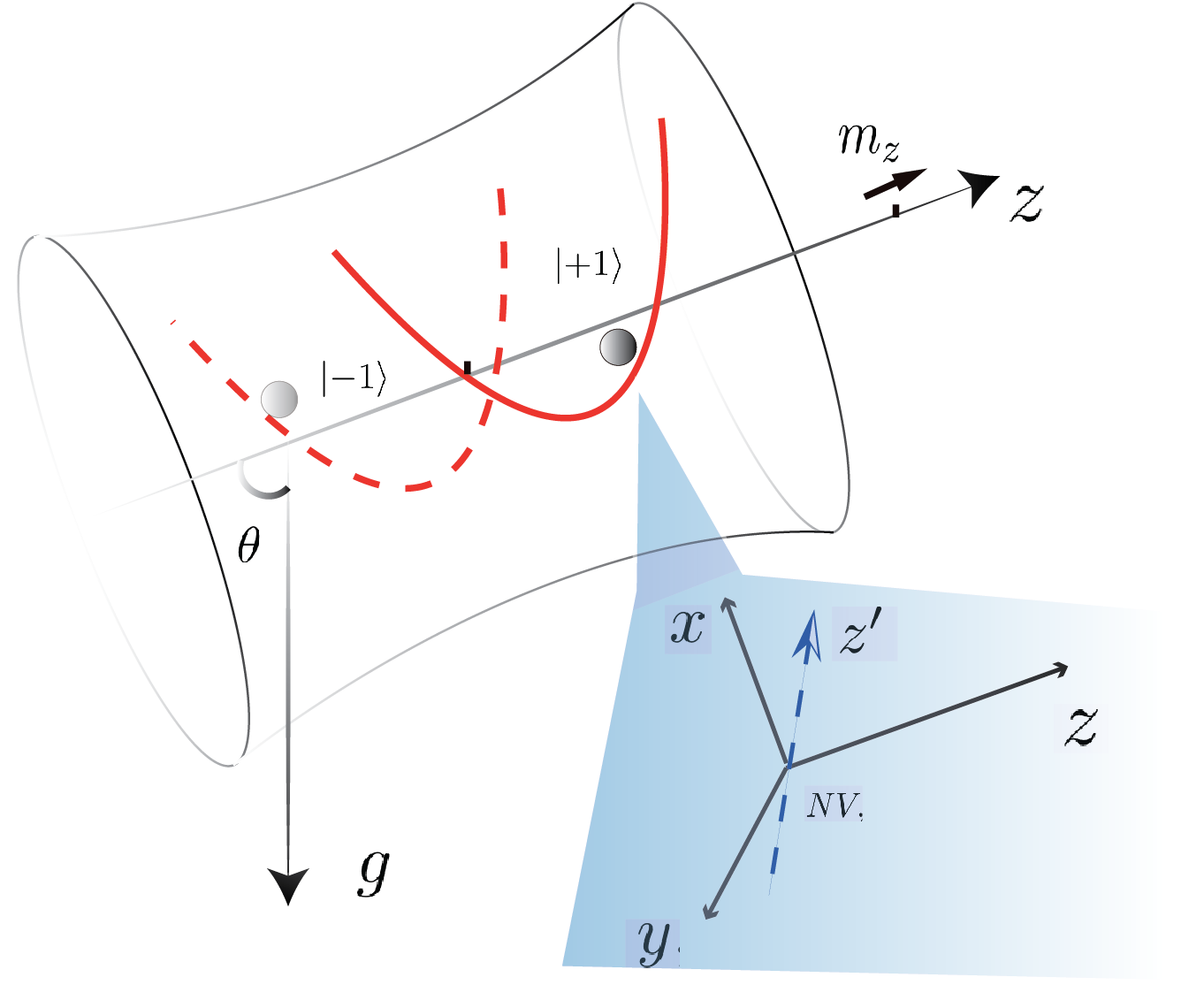}
 \caption{An optical trap holds a diamond bead with an NV center with both weakest confinement and spin quantization along the z axis. A magnetized sphere at $z_0$ produces spin-dependent shifts to the center of the harmonic well. An angle
 $\theta$ between the vertical and the $z$ axis places the centers of the wells corresponding to the $|+1\rangle$ and $|-1\rangle$ states in different gravitational potentials. Starting with an arbitrary coherent state, the CM of the bead oscillates as different coherent states in the centre-shifted, spin dependent well (red solid and dashed line), accumulating a relative gravitational phase difference due to the superpositions. At
 $t_0=2\pi/\omega_z$ this phase can be read from Ramsey measurements on spin. The blue shaped zone shows a generic orientation of the NV center' s axis $z^{\prime}$ with respect to magnetic direction $z$}
\label{setup}
\end{center}
\end{figure}
\begin{figure}[h]
\subfloat[\label{scheme}]{
       \includegraphics[width=0.45\textwidth]{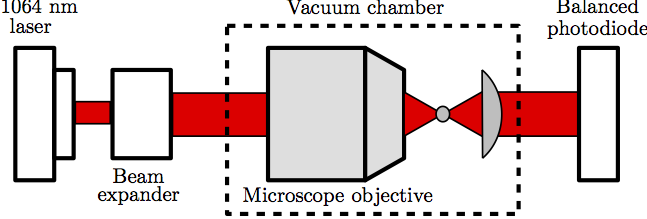}
       }
       \hfill
\subfloat[\label{psd}]{
       \includegraphics[width=0.45\textwidth]{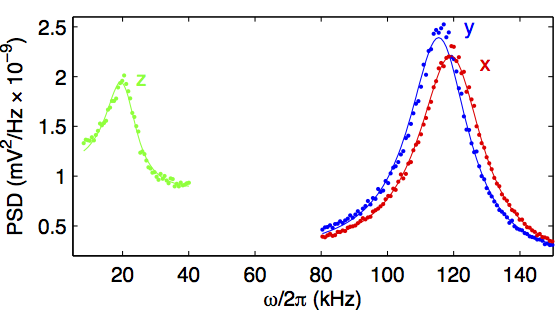}
       }
       \caption{(a) Experimental schematic of the optical dipole trap. A $1064$ nm laser beam is tightly focused by a high numerical aperture ($0.95$) objective. The polarization of the trapping light can be rotated by a half-wave plate. Scattered light from levitated nanodiamonds is collected by a lens and sent to a balanced photodiode in an interferometric scheme described in reference \cite{gieseler2012subkelvin}, providing a position dependent signal from the levitated nanodiamond. (b) Power spectral density (PSD) as a function of $\omega$ at approximately $10$ mB using $200$ mW of trapping power. Fourier transforming the position dependent signal yields the PSD of the trapped nanodiamond. The axial $z$ frequency has been scaled by a factor of $20$ for clarity.}
\label{fig2}
\end{figure}

As shown in Fig.\ref{setup}, the setup consists of a nano-scale diamond bead containing a single spin-1 NV center levitated by an optical tweezer in ultra-high vacuum. The motion of the bead is coupled to the $S=1$ spin of the NV center by means of  a static magnetic field gradient which can be generated by a magnetized sphere with a permanent dipole moment ${\bf m}=(0,0,m_z)$ oriented along the $z$ direction. Defining a reference frame such that the centers of the harmonic potential and the magnetized sphere be at $(0,0,0)$ and $(0,0,z_0)$ respectively, we can expand the magnetic field of the sphere around $(0,0,0)$, and get
\begin{equation}\label{bxy}
 B_x=-B_0\,x,\;\;B_y=-B_0\,y,\;\;B_z=\frac{\mu_0\,m_z}{2\pi\,|z_0|^3}+2B_0\,z,
\end{equation}
where $B_0=3\mu_0\,m_z/(4\pi z_0^5)$. Therefore the interaction between the
spin of the NV center and the vibrational motion can be described by the Hamiltonian
\begin{equation}\label{intham}
 H_{\rm int}=-\lambda\big[2\,S_z\,(c+c^\dag)-\sqrt{\frac{\omega_z}{\omega_x}}\,S_x\,(a+a^\dag)-\sqrt{\frac{\omega_z}{\omega_y}}\,S_y\,(b+b^\dag)\big],
\end{equation}
where
\begin{equation}\label{lambda}
\lambda=\frac{3\mu_0 m_z z_0}{4\pi |z_0|^5}\, g_{NV}\, \mu_B\,
\sqrt{\frac{\hbar}{2\, m \,\omega_z}},
\end{equation}
$m$ being the mass of the bead, $g_{NV}$ the Land\'e factor of the
NV center, $\mu_B$ the Bohr magneton and $a$, $b$, $c$ the annihilation operator of the oscillation in $x$, $y$, $z$ respectively. In this Section we will neglect the interaction between the spin and the $x$ and $y$ directions in Eq. (\ref{intham}), whose effect will be analyzed in the next Section, on the basis that $\omega_x,\omega_y\gg\omega_z$. 
In support of this approximation, in Fig. \ref{fig2} we present experimental data measured in our laboratory from a nanodiamond levitated in moderate levels of vacuum of approximately $10$ mB using $200$ mW of $1064$ nm trapping power. We measure $\frac{\omega_z}{\omega_x}\approx \frac{\omega_z}{\omega_y}\approx 0.18$. Due to asymmetry in the laser focus \cite{gieseler2012subkelvin}, oscillation frequencies in the radial directions are separated by approximately $\omega/2\pi=5$ kHz. The lower axial $z$ frequency arises from the smaller electric field gradient along the beam axis in comparison to $x$ and $y$. 
The $x$ or $y$ frequency is revealed to the balanced photodiode by rotating the polarization of the trapping light, while $z$ is measured as the nanodiamond traverses an intensity gradient along the beam axis, giving rise to a time varying scattering intensity.

Finally, we add the free Hamiltonian of the bead and of the NV center and the Hamiltonian describing the interaction between the bead and the Earth's gravitational field, i.e., $mgz\cos\theta$, where $\theta$ is the angle between the $z$ direction (see Fig.\ref{setup}), so that the total Hamiltonian of the system is given by 
\begin{equation}\label{Ham}
 H=D\,S_z^2+\hbar\omega_zc^\dag c-2(\lambda\,S_z-\Delta\lambda)(c+c^\dag),
\end{equation}
with
\begin{equation}\label{deltalambda}
 \Delta\lambda=\frac{1}{2}mg\cos\theta\sqrt{\frac{\hbar}{2m\omega_z}}.
\end{equation}
%
%
%
%
%
%
The Hamiltonian above represents a harmonic oscillator whose center depends on the eigenvalue of $S_z$. In its derivation we have also assumed that the Zeeman splitting of  $\left|+1\right>$
and $\left|-1\right>$ due to the zeroth order expansion of $B_z$ is cancelled by addition of a
uniform magnetic field along $z$.


It can be shown that, starting at $t=0$ from the state $ \left|\beta\right>\left|s_z\right>$, where $\left|\beta\right>$ is a coherent state of the center of mass (CM) quantized motion of the bead  in the
harmonic well centered at $z=0$ and $\left|s_z\right>$ is an eigenstate of the operator $S_z$ with eigenvalue $s_z=+1, 0, -1$, the system evolves at time $t$ to the state $\left|\beta (t, s_z)\right>\left|s_z\right>$  where
\begin{eqnarray}
\left|\beta (t, s_z)\right> =&
\mathrm{e}^{-\frac{i}{\hbar}(D-\hbar\omega_z u^2)t}
 \mathrm{e}^{iu^2\sin(\omega_z t)}\nonumber\\
 &
\times\left|(\beta-u)\mathrm{e}^{-i\omega_z t}+u\right>,
\label{dynamics}
\end{eqnarray}
and $u=2(s_z\,\lambda-\Delta\lambda)/\hbar\omega_z$.
It is worth noting that,  at time $t_0=2\pi/\omega_z$, the oscillator state returns to
its original coherent state $\beta$, for {\em any} $\beta$ and
$s_z$. One can take advantage of this feature  to show that spin measurements at $t_0$ will
be unaffected by any thermal randomness in the initial motional
state of the oscillator. 
%

\subsection{Detecting the gravitational field by Ramsey interferometry}
What we want to do now is to prepare the spin in a superposition of states, so that when the
center of mass of the bead undergoes a conditional displacement due to dynamics, the hybrid system evolves to a state which involves superpositions of $\left|\beta (t, s_z)\right>\left|+1\right>$ and $\left|\beta (t, s_z)\right>\left|-1\right>$. We then measure the spin at a special time $t_0$ at which the conditional displacements are again undone due to the natural dynamics so that the spin state becomes unentangled from the
vibrational one. In this way we can perform measurements on the
spin only, to reveal the different phases acquired by the
spin components, due to the evolution for the vibrational
state.

Preparing the system in the separable state
$\left|\beta\right>\left|s_z=0\right>$ and applying a microwave
pulse corresponding to the Hamiltonian
$H_{mw}=\hbar\Omega\left(\left|+1\right>\left<0\right|+\left|-1\right>\left<0\right|+\mathrm{h.c.}\right)$, with $\Omega$ much larger than any other coupling constant and for a pulse duration $t_p=\pi/(2\sqrt{2}\Omega)$, the
spin state becomes $ \left|\Psi(0)\right>=\left|\beta\right>\left(\frac{\left|+1\right>+\left|-1\right>}{\sqrt{2}}\right)$,
which we  will take as the initial state for the interaction under the Hamiltonian (\ref{Ham}). 
After the interaction time $t$, the state is 
\begin{equation}
 \left|\Psi(t)\right>=\left(\frac{\left|\beta(t,+1)\right>\left|+1\right>+\left|\beta(t,-1)\right>\left|-1\right>}{\sqrt{2}}\right),
\end{equation}
which is the superposition we intend to evidence.  From the
expressions of $\left|\beta(t,\pm 1)\right>$ in
Eq.(\ref{dynamics})  we can see that separated coherent states are involved in
the above superposition along with phases due to gravitational
potential, which will evidence the above
superposition. Up to a global phase factor, the state after an oscillation period
$t_0=2\pi/\omega_z$ is
\begin{equation}
 \left|\Psi(t_0)\right>=\left|\beta\right>\left(\frac{\left|+1\right>+\mathrm{e}^{i\Delta\phi_{\text{Grav}}}\left|-1\right>}{\sqrt{2}}\right),
\end{equation}
with
\begin{equation}\label{deltaphi}
 \Delta\phi_{\text{Grav}}=\frac{16\lambda\,\Delta\lambda}{\hbar^2\omega_z}\,t_0.
\end{equation}
The phase difference $\Delta\phi_{\text{Grav}}$ can be revealed by applying
$H_{mw}$ again. Indeed, after a time $t_p$, the population of the spin state with $S_z=0$ is:
\begin{equation}\label{population}
 P_0(t=t_0+t_p)=\cos^2\left(\frac{\Delta\phi_{\text{Grav}}}{2}\right),
\end{equation}
which gives a direct connection between the value of the phase
shift and spin population. As $\Delta\phi_{\text{Grav}} \propto g$ can never appear as a relative phase between spin states unless spatially separated states of CM were involved in the superposition $\left(\left|\beta(t,+1)\right>\left|+1\right>+\left|\beta(t,-1)\right>\left|-1\right>\right)/\sqrt{2}$ for $0< t < t_0$, the detection of
$\Delta\phi_{\text{Grav}}$ evidences such a superposition.

Finally, let us consider what happens if the initial state is the product of a thermal motional state $\rho_{th}$ and an eigenstate of the spin operator $S_z$. In fact, 
one can exploit the fact that the results
given above are independent of the amplitude $\beta$ and that any thermal
state $\rho_{th}$ of the motion can be written as $\rho_{th}= \int
\mathrm{d}^2\beta\,P_{th}(\beta)\left|\beta\right>\left<\beta\right|$, where $P_{th}$ is the
Glauber $P$ representation for the thermal state, to show
that, after the evolution over one oscillation period, the state of the system is
again factorizable and that the phase difference accumulated by
the spin states is not affected by the thermal motion. Basically, though a mixture of many
Schr\"odinger cats
$\left|\beta(t,+1)\right>\left|+1\right>+\left|\beta(t,-1)\right>\left|-1\right>$
is generated for $0< t < t_0$, the interference between the
components $\left|\beta(t,+1)\right>\left|+1\right>$ and
$\left|\beta(t,-1)\right>\left|-1\right>$ of the cat is
independent of $\beta$. This immunity of the interference to
thermal states hinges on the mass being trapped in a
harmonic potential. We assume that anharmonic effects of the
trapping potential will be avoided by
feedback cooling of our oscillator to mK temperatures \cite{li2011millikelvin,gieseler2012subkelvin}.

\section{The effect of the terms neglected: perturbative analysis}

%
%
%
%
%
%
%
%
%
%
%
%
%
%

In the previous discussion, we restricted our analysis to the one-dimensional case, i.e., we neglected the coupling between the spin and the motion of the trapped bead along the $x$ and $y$ directions on the basis that the confinement along such directions is much tighter than the trapping along the $z$ directions. Since the coupling terms neglected in Eq. (\ref{intham}) are proportional to
\begin{equation}
 \gamma_{x,y}=\sqrt{\frac{\omega_z}{\omega_{x,y}}}
\end{equation}
it is worth analyzing the effect of such terms on the dynamics predicted in the previous section. In the following we will use a perturbative approach, the main scope of the treatment being the effect on the interference fringes corresponding to the gravity-induced phase difference in Eq. (\ref{deltaphi}). In order to do so, we first define the appropriate zeroth-order Hamiltonian as 
\begin{eqnarray}\label{Ham0}
 &&H_0= D\,S_z^2+\hbar\omega_xa^\dag a+\hbar\omega_yb^\dag b+\hbar\omega_zc^\dag c\nonumber\\
 &&+2\Delta\lambda_x(a+a^\dag)+2\Delta\lambda_y(b+b^\dag)+2\Delta\lambda(c+c^\dag)
\end{eqnarray}
which differs from the Hamiltonian in Eq. (\ref{Ham}) for the fact that we have added the quantized motion along the directions $x$ and $y$, representing by the annihilation operators $a$ and $b$, respectively. Subsequently the extra gravitational terms due to these transverse oscillation has been included through the constants:
\begin{equation}
 \Delta\lambda_{x,y}=\frac{1}{2}mg\cos\theta_{x,y}\sqrt{\frac{\hbar}{2m\omega_{x,y}}}
\end{equation}
where $\theta_x$ and $\theta_y$ are the angles between the direction of the gravitational acceleration and the $x$ and $y$ directions respectively.


The evolution of the system with the full Hamiltonian
\begin{eqnarray}\label{Hfull}
 H&=&H_0+V_x+V_y\nonumber\\
 V_x&=&\lambda\gamma_x S_x\,(a+a^\dag)\,,\,V_y=\lambda\gamma_y S_y\,(b+b^\dag)\,,
\end{eqnarray}
can be obtained by treating the terms $V_x$ and $V_y$ perturbatively with respect to the Hamiltonian (\ref{Ham0}), whose eigenstates are products of eigenstates of $S_z$ and displaced number states \cite{DisplacedNumber}:
\begin{equation}\label{Displaced-NS}
 \Ket{E_{{\bf n},s_z}^{(0)}} = D({\bf \alpha})\Ket{n_z, n_y, n_z}\otimes\Ket{ s_z},
\end{equation}
correspondingly the unperturbed eigenvalues reads:
\begin{equation}
E_{\bf n, s_z}^{(0)}=\hbar\sum_{i = x, y, z}\omega_i n_i + Ds_z^2 - 4\frac{(-s_z\lambda + \Delta \lambda_z)^2}{\hbar\omega_z}.
\end{equation}\label{Displaced-NS-eigenvalues}
In Eq. (\ref{Displaced-NS}), we have ${\bf n} = (n_x, n_y, n_z)$ and $D({\bf \alpha})$ is a three-dimensional displacement operator, i.e., $D({\bf \alpha})=D_x(\alpha_x)D_y(\alpha_y)D_z(\alpha_z)$, with
\begin{equation}
  \alpha_z=\frac{-2\lambda s_z+2\Delta\lambda_z}{\hbar\omega_z},
  \alpha_x=\frac{2\Delta\lambda_x}{\hbar\omega_x},
  \alpha_y=\frac{2\Delta\lambda_y}{\hbar\omega_y}.
\end{equation}

The first-order correction to the energy due to $V_x, V_y$ is zero since
\begin{equation}
\Bra{s_z}V_{x,y}\Ket{s_z} = 0.
\end{equation}	
The first nontrivial deviation from the underperturbed dynamics can be obtained by correcting the energies up to second order and, accordingly, the eigenstates up to first order. The corrected energy eigenstates and eigenvalues are therefore given by
\begin{equation}\label{EigenvecSec}
  \Ket{E_{\bf n, s_z}^{(2)}}=Z_n^{-1}(\Ket{E_{\bf n, s_z}^{(0)}}+\sum_{k\neq n, s_z'} \frac{H^{\prime}_{n.s_z;k,s_z'}}{E_{\bf n, s_z}^{(0)}-E_{\bf k, s_z'}^{(0)}}
  \Ket{E_{\bf k, s_z'}^{(0)}})\\
\end{equation}
and
\begin{equation}
  E_{\bf n, s_z}^{(1)}=E_{\bf n, s_z}^{(0)}+\sum_{k\neq n, s_z'}\frac{\left|H^{\prime}_{n.s_z;k,s_z'}\right|^2}{E_{\bf n, s_z}^{(0)}-E_{\bf k, s_z'}^{(0)}}
\end{equation}
where:
\begin{equation}
 H^{\prime}_{n,s_z;k,s_z'}=\Bra{E_{\bf n, s_z}^{(0)}}(V_x+V_y)\Ket{E_{\bf k, s_z'}^{(0)}},
\end{equation}
and $Z_n^{-1}$ is an appropriate normalization factor. 

For any initial state, the perturbative dynamics at any time $t$ can be obtained by expanding it into the basis given by Eq. (\ref{EigenvecSec}) and multiplying each term in the expansion by the corresponding exponential factor $\exp (-\frac{i}{\hbar} E_{\bf n, s_z}^{(1)} t)$.
In particular we have calculated the evolution of our system from the same initial state as in Sec.II. As a result, the motional and spin state will be entangled at every time instant $t>0$. We can quantify how much the perturbed evolution deviates from the unpertutbed one by computing the fidelity $F=\left|\left<\Psi^{(2)}(t_0)\right.\left|\Psi^{(0)}(t_0)\right>\right|$ between the perturbed state $\Ket{\Psi^{(2)}(t_0)}$ and unperturbed state $\Ket{\Psi^{(0)}(t_0)}$ after one oscillation period. We have computed fidelities for a very large range of parameters, with $\lambda$ ranging from zero to $0.1$ and $\gamma_x$ and $\gamma_y$ ranging from zero to $0.4$: this choice includes also the realistic experimental conditions of our proposal,  where $\omega_x=\omega_y=10\omega_z$, corresponding to $\gamma_x=\gamma_y\simeq 0.32$.
We have found that the fidelity decreases with increasing values of either $\lambda$, $\gamma_x$ and $\gamma_y$, but only by a very small amount in the range considered: in all cases the value of the fidelity stays above $99\%$, which assures that our treatment in Sec. II is accurate enough. This is somehow surprising, but the reason relies on the fact that the first-order correction in the energies is exactly zero due to the selection rules, so that the first significant correction to the energies is second order in $\gamma_x$ and $\gamma_y$, so that even relatively large values of these parameters have little effect on the evolution of the spin-diamond system.

These results remain true when the initial state is thermal. We have indeed found that the fidelity is still very close to unity for coherent states whose average thermal occupation  number is up to $600$. This corresponds to robustness of our schemes up to temperatures of the order of 1mK, so that only feedback cooling is sufficient for our scheme to work \cite{li2011millikelvin}.

\section{The effect of the random orientation}

Once the nanodiamond is trapped in the potential, it is very likely that the orientation of the NV center quantization axis, with respect to which the splitting $D$ is computed, is randomly orientated with respect to the trap axis. 
Measuring the optically-detected magnetic resonance spectrum of the NV
electron spin in an applied magnetic field will reveal the orientation of
the NV center with respect to the applied magnetic field and hence the trap
axis. This orientation could be controlled in x-y plane by adjusting the linear
polarization of the trapping light because the nanodiamonds are not
spherical. Using rod-shaped diamonds would increase this control \cite{shelton2005nonlinear}.
Alternatively, the birefringence of diamond \cite{lang1967causes} might be used to control the orientation with circularly polarized light \cite{arita2013laser}. 

In this section, without introducing extra measurements and manipulation on the NV's orientation, we will evaluate the systematic errors introduced by the misalignment between the axes of NV center and the trap. We will show that the visibility of the interferometry will reduce to some degree when these two axes are perpendicular, but in most cases we can get a good resolution of the interference fringes. We will essentially use the same perturbative methods as in the previous section. 

Assuming we can neglect the coupling with the $x$ and $y$ directions, the Hamiltonian we will use is
\begin{equation}
\begin{aligned}
H=&DS_z^2+\hbar\omega_zc^{\dagger}c+2\Delta\lambda(c^{\dagger}+c)\\
&-2\lambda (c_xS_x+c_yS_y+c_zS_z)(c^{\dagger}+c),\\
\label{P_O_H_1}
\end{aligned}
\end{equation}
where $c_x,c_y$ and $c_z$ are the direction cosines of the NV symmetry axis $z'$ with respect to the trapping axis $z$ and whose values are obtained by a rotation transformation between these two axes, bounded by the condition $c_x^2 + c_y^2  + c_z ^2 = 1$. Generally there is no exact analytical solution for this Hamiltonian, so we divide it into two parts by $H=H_0+H_I$, where 
\begin{equation}
H_0=DS_z^2+\hbar\omega_zc^{\dagger}c+2\Delta\lambda(c^{\dagger}+c)-2\lambda c_zS_z(c^{\dagger}+c),
\end{equation}
for which we can solve analytically, and 
\begin{equation}
H_I=-2\lambda(c_xS_x+c_yS_y)(c^{\dagger}+c),
\end{equation}
which will be treated perturbatively. The time evolution of any state under Hamiltonian (\ref{P_O_H_1}) and the effect on the Ramsey scheme afterwards can be calculated numerically by means of a  perturbative expansion to the second order, and based on numerical results we reconstruct the fringes of spin zero state population over a range of angles $\theta$ between the trapping axis $z$ and the direction of the gravitational  acceleration.

We can investigate the dependence on the orientation  of the NV center by changing the parameters $c_x$ and $c_y$. Since the system is rotationally invariant along the trapping axis, we simplify the simulation by taking $c_y = 0$ and consider $c_x$ varying from 0 (aligned case) to 1 (perpendicular case).

Fig. 3 shows how the interference fringes, given by the population $P_0(t_0+t_p)$ of the state $\Ket{s_z=0}$, after the last Ramsey pulse has been sent, as a function of the tilting angle 
$\theta$ and of the direction cosine $c_x$ ($c_x=0$ corresponds to the NV center being parallel to the trapping axis, while $c_x=1$ corresponds to the case in which the NV center is orthogonal to it). Fig. 3 compares the interference fringes in the case of perfect alignment with the fringes in the orthogonal case. The interferometry fringes turn out to be qualitatively robust against the misalignment, i.e., the visibility of the fringes does not change over quite a long range of values of $c_x$ when the NV axis is deviating from the paraxial case, and only show a limited reduction at the extreme  case. We can therefore deduce that, if we want to perform a proof-of-principle experiment to show the existence of the gravity-dependent phase factors in the evolution of the components of the spin state, we do not have to worry too much about correcting the orientation of the NV center with respect to the trapping axis.

\begin{figure}[h]
\centering
\includegraphics[width=9cm]{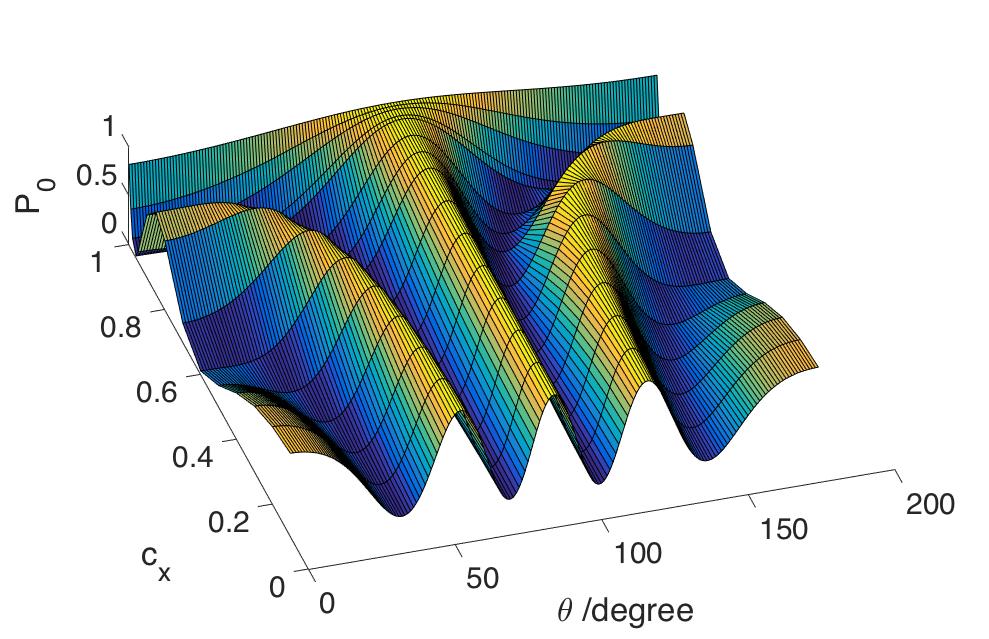}
\caption{Fringes of spin zero population $P_0$ as a function of the orientation $\theta$ of the trapping axis, z, with respect to the direction of the gravitational acceleration, and of the direction cosine $c_x=0$ corresponds to the NV center being parallel to the trapping axis, while $c_x=1$ corresponds to the case in which the NV center is orthogonal to it. The initial motional state has been taken equal to the vacuum state of the quantum oscillator. The other parameters are such that  $\lambda = 0.01$ J and $\Delta\lambda /\cos\theta = 11.9$ J.}
\label{NV-config1}
\end{figure}


\begin{figure}[h]
\centering
\includegraphics[width=9cm]{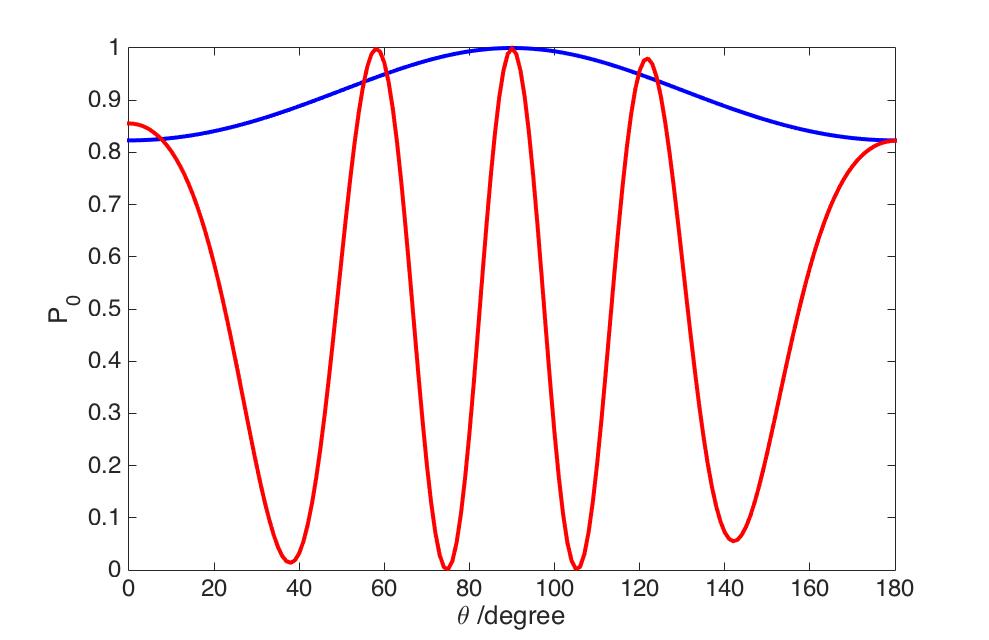}
\caption{Details on how the fringes change when the spin is orthogonal to the trapping axis: the visibility reduces but does not vanish at the perpendicular case.}
\label{NV-config1}
\end{figure}

These results in Fig. 3 and Fig. 4 have been obtained with  $\lambda=0.01$ J and $\Delta\lambda / \cos \theta=10$ J, and for a motional initial state corresponding to the vacuum state of the quantum oscillator. In order to investigate the effect of an initial thermal distribution we must carry the simulation starting from a generic initial coherent state for the oscillator. For computational purposes we have simulated the dynamics with the QuTip software \cite{johansson2013qutip}, which assures faster convergence than our perturbative approach, and  computed the evolution of the system with a generic coherent state. Up to an average thermal occupation number of CM motion $\braket{N}=600$  we have not noted any visible changes in the behavior of our system, which assures us that the initial thermal distribution should not affect our predictions at least up to temperature of the order of 1mK.


\section{Discussion and conclusive remarks}

We will now give the experimental
parameters necessary to obtain a good visibility of
the interferometric fringes in a setup in which we
are allowed to vary the angle $\theta$. As realistic values for the setup, we
consider $\omega_z \sim 100$ kHz and diamond spheres whose radius
$R \sim 100$ nm, so that, considering the density
$3500\,\mathrm{kg}/\mathrm{m}^3$ for diamond, the
corresponding mass is $\sim 1.25 \times 10^{-17}$ kg. A
good visibility of interferometry fringes in the population in Eq.
(\ref{population}) is given for
$K=8\lambda\,\Delta\lambda\,t_0/(\hbar^2\omega_z\cos\theta)\sim10$,
which makes the value of the population change completely from $0$
to $1$ when $\theta$ varies between $\pi/2-\pi/20$ and $\pi/2$
(the $z$ axis is horizontal for $\theta=\pi/2$) (see red line in Fig. 3). Assuming that the
magnetic field in Eq. (\ref{bxy}) is generated by a magnetized
sphere with radius $r_0=40\,\mu$m and magnetization $M=1.5 \times
10^{6}\, \mathrm{A}/\mathrm{m}$ (typical for commercial
magnets), and $z_0=120\,\mu$m, we get, according to $m_z=M\cdot
(4\pi/3)r_0^3$ and Eqs. (\ref{lambda}) and (\ref{deltalambda}) we have a magnetic gradient of $\partial B/\partial z \sim 10^7 T/m$ and consequently the desired value of K. All these values are achievable experimentally and correspond to the values of $\lambda$ and $\Delta\lambda$ used to obtain Figs. 2 and 3. 

With these values, the time necessary to have a complete oscillation of the center of mass is of the order of $t_0=50\mu$s. This must be compared with the typical spin dephasing times of NV centers in nanodiamond. NV centers in isotopically-purified bulk diamond can have electron spin dephasing times $T_2>1$ms measured with a spin echo sequence
\cite{balasubramanian2009ultralong}, but such long
times have not been found for nanodiamonds. The longest times have been
achieved by making nanodiamonds from pure bulk material with a low
concentration of spin defects: nitrogen impurities must be reduced and
ideally also $^{13}$C. This material can then be milled \cite{knowles2014observing} or preferably etched with reactive ion etching (RIE) to form nanoparticles. Nanodiamond pillars with
300-500nm diameter, made with RIE have shown a spin-echo $T_2$ time of over
300 $\mu$s \cite{andrich2014engineered}. Pillars with 50
nm diameter and 150 nm length have achieved a spin-echo $T_2$ time of 79
$\mu$s \cite{trusheim2013scalable}. By means of
appropriate decoupling techniques, and by slightly changing our Ramsey
scheme in order to accumulate the gravitational phase over more than one
cycle, such a dephasing time can be made larger. With $T_2 = 79 \mu$s we
would get perfect visibility of the interference fringes.

The robustness to the temperatures up to 1mK, also shows that our proposal is promising for an experimental realization. As we have shown, both the terms neglected in obtaining our results in Sec. II and the misalignment terms are not a problem if we want to perform an experiment aimed at obtaining interference fringes demostrating the existence of gravity-induced phases in the spin state.

For other applications of our scheme, such as in metrology, we will need a perfect quantitative agreement between our predictions and the data obtained. In this case, we will either have to measure the exact value of $c_x$ in order to know to which slice of Fig. 3 our fringes correspond, or we will have to correct for misalignment. This is currently experimentally achievable, for example by the methods shown in Ref. \cite{arita2013laser}.

Finally let us stress that our perturbative approach can be useful to study the effect of similar spurious coupling terms in experimental proposals involving the coupling of the motion of nanoparticles with spin degrees of freedom \cite{albrecht2014testing,yin2013large,rabl2009strong}.

In this paper we have studied the dynamics of an NV center in a harmonically trapped nanodiamond, considering the effect of unwanted coupling between the spin and the directions orthogonal to the direction where the conditional displacement takes place, inducing a gravity-dependent phase which can be detected by spin measurements only. 
We have shown that a perturbative treatment of the additional terms in the Hamiltonian proves that, for the experimental parameters characterizing the setup proposed, the fidelity between the realistic and the unperturbed state is always above $99\%$. This assures that the experiment should confirm the one-dimensional theory, which allows for a very clear physical interpretation of the results. Moreover, also the robustness with respect to thermal fluctuations in the initial state is retained, which is the most important feature of our proposal, since no ground state cooling is required, only some feedback cooling to mK temperatures in order to guarantee that no anharmonic effects come into play in the dynamics. The same perturbative approach can be applied to treat the misalignment between the NV center and the trapping axis. In the case of misalignment, the quantitative agreement with the one-dimensional model is lost, but the qualitative behavior of the interference fringes stays valid up to very high deviations from the perfectly aligned case. Therefore, a proof-of-principle experiment aimed at showing the existence of Ramsey interference fringes induced by gravity can be performed without caring too much about the alignment between the NV center and the trapping axis. 
\\



\section*{Acknowledgments}
We acknowledge the EPSRC grant EP/J014664/1. This work was also supported by EPSRC as part of the UK Hub in Networked Quantum
Information Technologies (NQIT), grant EP/M013243/1. GWM is supported by the Royal Society. C. Wan is sponsored by Imperial CSC scholarship.

\end{document}